\documentclass{pasj00}

\begin{document}
\SetRunningHead{Y. Takeda}{Spectroscopic Study of Solar Twins:
HIP~56948, HIP~79672, and HIP~100963}
\Received{2008/12/02}
\Accepted{2009/01/07}

\title{High-Dispersion Spectroscopic Study of Solar Twins: \\
HIP~56948, HIP~79672, and HIP~100963
\thanks{Based on data collected at Subaru Telescope, which is operated
by the National Astronomical Observatory of Japan. 
The electronic tables E1 and E2 will be available 
at the E-PASJ web site upon publication, while they are provisionally 
placed at the WWW site of
$\langle$http://optik2.mtk.nao.ac.jp/$\ \widetilde{ }\ $takeda/solartwins/$\rangle$.}
}

%

\author{Yoichi \textsc{Takeda}}
\affil{National Astronomical Observatory, 2-21-1 Osawa, Mitaka, Tokyo 181-8588}
\email{takeda.yoichi@nao.ac.jp}
\and
\author{Akito \textsc{Tajitsu}}
\affil{Subaru Telescope, 650 North A'ohoku Place, Hilo, Hawaii 96720, U.S.A.}
\email{tajitsu@subaru.naoj.org}

\maketitle

\begin{abstract}
An intensive spectroscopic study was performed for three 
representative solar twins (HIP~56948, HIP~79672, and HIP~100963)
as well as for the Sun (Moon; reference standard), with an intention
of (1) quantitatively discussing the relative-to-Sun 
similarities based on the precisely established differential 
parameters and (2) investigating the reason causing the
Li abundance differences despite their similarities.
It was concluded that HIP~56948 most resembles the Sun in 
every respect including the Li abundance (though not perfectly
similar) among the three and deserves the name of ``closest-ever 
solar twin'', while HIP~79672 and HIP~100963 have somewhat higher 
effective temperature and appreciably higher surface Li composition.
While there is an indication of Li being rotation-dependent 
because the projected rotation in HIP~56948 
(and the Sun) is slightly lower than the other two, the rotational 
difference alone does not seem to be so large as to efficiently 
produce the marked change in Li. Rather, this may be more likely
to be attributed (at least partly) to the slight difference in 
$T_{\rm eff}$ via some $T_{\rm eff}$-sensitive Li-controlling mechanism.
Since the abundance of beryllium was found to be essentially solar 
for all stars irrespective of Li, any physical process causing 
the Li diversity should work only on Li without affecting Be.

\end{abstract}

%


\section{Introduction}

Can we find such a star that indiscernibly resembles our Sun
in every respect? This ``solar twin\footnote{We use 
the term ``solar twin'' for those special solar-type stars 
which have particularly high similarity to the Sun 
with respect to spectra as well as stellar parameters.
See Appendix A of Takeda et al. (2007) and the references therein
for the literature concerning this theme.} survey'', 
an ever-attracting subject for stellar astronomers, has made 
significant progress since 1990s, thanks to the improvement 
in the precision of stellar parameter determinations. 

Since Porto de Mello and da Silva (1997) reported the remarkable
similarity of HIP~79672 (= 18~Sco = HR~6060 = HD~146233; $V = 5.50$)
to the Sun, this star has maintained the status of best solar twin
candidate almost for a decade (see also Soubiran \& Triaud 2004). 
In the meantime, by using a numerical technique developed 
by Takeda (2005; hereinafter referred to as Paper I)
for establishing the parameter differences between two similar
stars with high precision, Takeda et al. (2007; hereinafter 
Paper II) conducted a comprehensive study of solar analog stars
and found that HIP~100963 (= HD~195934; $V = 7.09$) is an equally 
good (or even better) solar twin as HIP~79672.

Yet, there is one concern. While these HIP~79672 and HIP~100963 are 
certainly very similar to the Sun in terms of the stellar parameters 
and the general appearance of the spectra, one marked dissimilarity 
exists in a particular part of the spectrum: the strength of Li line 
at 6707.8~$\rm\AA$ in these two stars is appreciably stronger compared 
to the solar case (cf. figure 3 in Soubiran \& Triaud 2004 and 
figure A.2 in Paper II). This decisive difference in the surface
Li abundance is actually ``a fly in the ointment,'' which makes us 
somewhat hesitate to regard them as ``real'' solar twins.

Interestingly, however, Mel\'{e}ndez and Ram\'{\i}rez (2007)
recently reported that HIP~56948 (= HD~101364; $V=8.70$) appears 
to be an ideal solar twin in the sense that it has essentially 
solar parameters and the low Li abundance similarly to the Sun.
If this is confirmed, this star may deserve being called as
a genuine solar twin. It would thus be worth carrying out
an independent check analysis in order to ascertain whether HIP~56948 
really resembles our Sun on every point including the Li abundance.

Another related subject of interest is the cause of such a difference
in the Li abundance among these superficially very similar solar twins. 
It was concluded in Paper II based on the analysis of 118 solar-analog 
dwarfs around early-G type that the surface Li abundance is closely
correlated with the macroscopic line-broadening parameter
(comprising macroturbulence plus rotation); i.e., the surface Li tends 
to be higher (i.e, less depleted) as the line-width becomes broader
(cf. figure 13 therein). Since the macroturbulence (due to the granular 
motion of stellar convection origin) is unlikely to differ much among 
similar solar-type stars, this observational fact suggests that the 
most decisive factor controlling the surface Li abundance is the 
stellar rotation or the angular momentum (i.e., faster rotation tends
to suppress the envelope mixing and leads to less depletion of Li). 
Then, is the distinction between the Li-strong (HIP~79672 and HIP~100963) 
and Li-weak (HIP~56948 and naturally the Sun itself) solar twins 
simply caused by the difference in the rotational velocity?
This point should be checked by careful determinations of the 
projected rotational velocities of these stars.

Besides, we should also pay attention to two other related viewpoints 
in connection with this ``rotation--mixing--surface Li''
relationship. The first is the stellar activity which tends to 
diminish/enhance as the rotation becomes slower/faster. 
If the rotation is the key factor affecting the surface Li, does the 
Li-strong solar twins show higher activity than Li-weak ones?
The second is the surface abundance of beryllium, which is destroyed
when conveyed into the hot stellar interior by envelope mixing similarly 
to lithium at temperature of $T \sim 3.5 \times 10^{6}$~K (higher than 
the case of Li which is burned at $T \sim 2.5 \times 10^{6}$~K).
It is interesting to see whether any difference is observed in the 
surface abundance of Be between the Li-strong and Li-weak groups, 
which may provide us with an observational constraint on the origin 
of Li discrepancies among these solar twins

Motivated by these considerations, we decided to conduct an 
intensive spectroscopic study for these representative solar twins 
(HIP~56948, HIP~79672, and HIP~100963 along with the Sun/Moon as 
the comparison standard) based on the high-dispersion spectra 
obtained by the Subaru Telescope with HDS, in order to (1) quantitatively
discuss the relative-to-Sun similarities of these three stars while 
precisely establishing their ``star$-$Sun'' differential parameters
by applying the method of Paper I and to (2) investigate the 
reason/mechanism causing the difference between the Li-strong and 
Li-weak groups by examining the rotational velocity, the degree of 
stellar activity, and the Li as well as Be abundance. This is the 
purpose of this study.

\section{Observational Data}

The observations of HIP~56948, HIP~79672, HIP~100963 and the Moon
(substitute for the Sun)  were carried out in the night of 2008 
June 15 (Hawaii Standard Time) by using the High Dispersion Spectrograph 
(HDS; Noguchi et al. 2002) placed at the Nasmyth platform of the 
8.2-m Subaru Telescope, which can record high-dispersion spectra 
covering a wavelength portion of $\sim 1600 \rm\AA$ (blue cross disperser)
or $\sim 2600 \rm\AA$ (red cross disperser) with two CCDs 
of 2K$\times$4K pixels at a time. 

In order to cover the wide wavelength range from near-UV ($\sim 3000 \rm\AA$)
to red ($\sim 7000 \rm\AA$), each star was observed at two different
wavelength settings (standard Ub with blue cross disperser for 
$\sim$~3000--4500~$\rm\AA$ and standard Yc with red cross disperser 
for $\sim$~4400--7000~$\rm\AA$).  
With the slit width set at $0.''4$ (200 $\mu$m) and no on-chip binning of
pixels, the resolving power of the obtained spectra is $R \simeq 90000$.
The integrated exposure times for each star at Ub/Yc settings are 
64~min/32~min, 18~min/7~min, 32~min/16~min, and 2~min/0.5~min
for HIP~56948, HIP~79672, HIP~100963 and the Moon, respectively.

The reduction of the spectra (bias subtraction, flat-fielding, 
scattered-light subtraction, spectrum extraction, wavelength calibration,
continuum normalization) was performed by using the ``echelle'' package of 
the software IRAF\footnote{IRAF is distributed
    by the National Optical Astronomy Observatories,
    which is operated by the Association of Universities for Research
    in Astronomy, Inc. under cooperative agreement with
    the National Science Foundation.} 
in a standard manner. 
The estimated S/N ratios at each of the wavelengths calculated as 
the square root of the resulting photoelectron counts (ADN $\times$ $gain$)
are graphically depicted in figure 1. We can see from this figure
that sufficiently high S/N ratios of $\sim$~500--1000 are achieved
in the most sensitive red region, though this value is considerably
reduced even by a factor of $\sim$~10 at $\sim$~3100~$\rm\AA$ of 
near-UV where Be~{\sc ii} lines are located.

\setcounter{figure}{0}
\begin{figure}
  \begin{center}
    \FigureFile(70mm,100mm){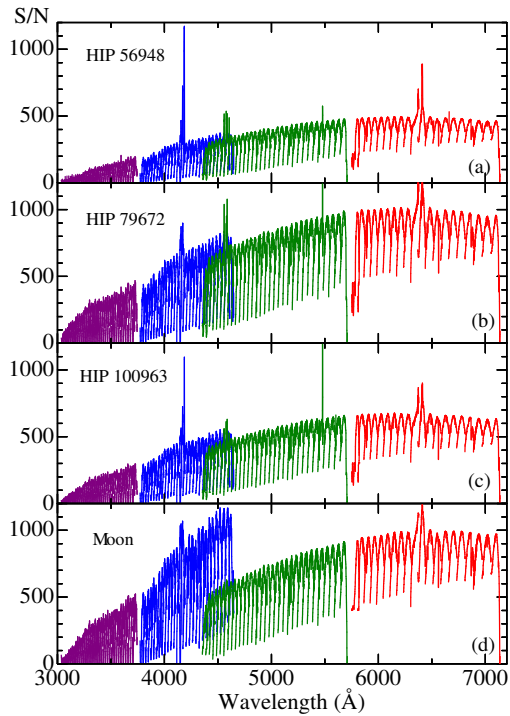}
  \end{center}
\caption{Distribution of S/N ratios (estimated as the 
square root of photoelectron counts) of the spectra used for
this study, which are divided into four wavelength regions: 
$\sim$~3000--3700~$\rm\AA$ (blue CCD) and $\sim$~3700--4600~$\rm\AA$ 
(red CCD) in the Ub setting; $\sim$~4400--5700~$\rm\AA$ (blue CCD) 
and $\sim$~5700--7000~$\rm\AA$ (red CCD) in the Yc setting. 
The absence of data at $\sim 3700\rm\AA$ and $\sim 5700\rm\AA$
corresponds to the joint of mosaicked CCDs. Spurious 
spikes seen at several wavelengths are due to bad columns
of CCDs. (a) HIP~56948, (b) HIP~79672, (c) HIP~100963, 
and (d) Moon.
}
\end{figure}

\section{Parameter Determination}

\subsection{Standard Stellar Parameters}

Following the procedure described in subsubsection 3.1.1
of Paper II, we measured the equivalent widths (EW) of Fe~{\sc i}
and Fe~{\sc ii} lines on the ``Yc setting'' spectra covering 
$\sim$~4400--7000~$\rm\AA$. 
Based on these EW values, the four standard\footnote{We use the notation of
``standard'' here, which means that these have characters of ``absolute'' 
parameters in the usual sense, in order to discriminate them from 
the ``differential'' parameters (relative to the Sun) presented in subsection 3.5.} 
atmospheric parameters 
[$T_{\rm eff}^{\rm std}$ (effective temperature), $\log g$ (surface gravity), 
$v_{\rm t}$ (microturbulent velocity dispersion), and \{Fe/H\}$^{\rm std} 
[\equiv A_{\rm Fe}^{\rm std} - 7.50]$ (Fe abundance\footnote{We intentionally 
expressed this quantity as \{Fe/H\}, in order to clarify that it is still
an absolute quantity (i.e., essentially equivalent to $A_{\rm Fe}^{\rm std}$)
and should be distinguished from the strictly differential metallicity
relative to the Sun, [Fe/H]  ($\equiv A_{\rm Fe}^{\rm star} - A_{\rm Fe}^{\rm sun}$), 
which is also denoted as $\Delta A_{\rm Fe}$ in subsection 3.5.}
)], which are necessary for constructing model atmospheres, were spectroscopically 
derived by using the TGVIT program (Takeda et al. 2005; cf. section 2 therein). 
This method is based on the principle searching for the most optimum solution
in the 3-dimensional ($T_{\rm eff}$, $\log g$, $v_{\rm t}$) space
such that simultaneously satisfying the three requirements of (i) 
the excitation equilibrium, (ii) the ionization equilibrium, and 
(iii) the EW-independence of the abundances (cf. Takeda, Ohkubo, \& Sadakane 2002). 
The detailed EW data and the Fe abundances corresponding to the 
final parameters for each star are presented in electronic table E1.
The results are summarized in table 1, where the related stellar 
parameters ($L$, $M$, and $age$) evaluated as in subsection 3.3 of 
Paper II are also given.

Comparing the present EW data measured from the Subaru/HDS spectra
($R \simeq 90000$, S/N $\sim$ 500--1000) with those from OAO/HIDES
spectra in Paper II and Paper I ($R \simeq 70000$, S/N $\sim$ 200--600),
we confirmed a general consistency as shown in figure 2. However,
a close inspection revealed a slight systematic difference in the sense 
that EW(Subaru) tends to be by $\sim$~1--2\% smaller than EW(OAO).
According to this delicate systematic change, marginal differences
are seen in the present results of such absolute parameters when 
compared to those in these previous papers; e.g., for the case of 
the Sun/Moon, $T_{\rm eff}$ and \{Fe/H\} have been lowered by 
$\sim 30$ K and $\sim 0.03$~dex, respectively, though these changes
should not matter in the differential analysis (subsection 3.5).

\setcounter{figure}{1}
\begin{figure}
  \begin{center}
    \FigureFile(70mm,70mm){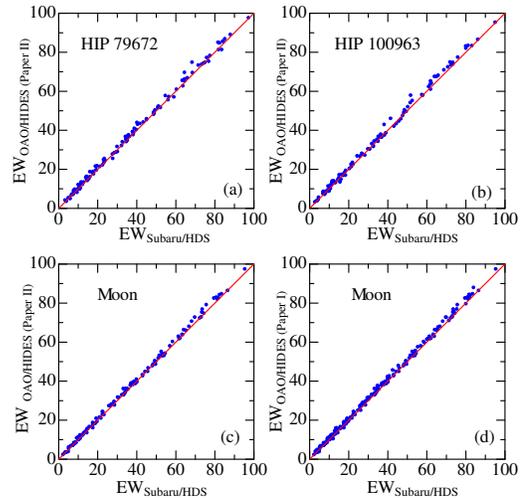}
  \end{center}
\caption{Comparison of the equivalent widths of Fe~{\sc i} 
and Fe~{\sc ii} lines measured on the Subaru/HDS spectra 
for the determinations of atmospheric parameters in this 
study (abscissa) with those used in the previous investigations 
based on the OAO/HIDES spectra (ordinate).
(a) HIP~79672 (Paper II), (b) HIP~100963 (Paper II), 
(c) Moon (Paper II), and (d) Moon (Paper I).
}
\end{figure}

\subsection{Rotational Velocity}

In order to evaluate the stellar projected rotational velocity
($v_{\rm e} \sin i$), we determined the total macrobroadening
parameter, $v_{\rm M}$, which is the $e$-folding width of 
the Gaussian macrobroadening function, 
$f_{\rm M} (v) \propto \exp [-(v/v_{\rm M})^{2}]$,
by way of the line-profile fitting as done in subsection 4.2 
of Paper II. Unlike the previous case (where the fitting was applied
to the spectrum portion at the 6080--6089~$\rm\AA$ region), however, 
we performed the fitting analysis to each of the ``individual''
Fe~{\sc i} and Fe~{\sc ii} lines (the same lines as used for 
the EW measurements in subsection 3.1) as Takeda (1995) did
for the solar flux spectrum, because the macroturbulence 
(to be subtracted from the total macrobroadening) is 
considered to be different from line to line because of its 
depth-dependence (cf. Takeda 1995).

We used the line-broadening model adopted by Takeda et al. (2008).
That is, the total macrobroadening function, $f_{\rm M} (v)$, 
is assumed to be the convolution of three Gaussian component 
functions $f_{\alpha} \propto \exp [-(v/v_{\alpha})^{2}]$, 
where $\alpha$ is any of ``ip'' (instrumental profile), 
``rt'' (rotation), and ``mt'' (macroturbulence); i.e.,
\begin{equation}
v_{\rm M}^{2} = v_{\rm ip}^{2}+v_{\rm rt}^{2}+v_{\rm mt}^{2}
    \; \; \; (= v_{\rm ip}^{2} + v_{\rm r+m}^{2}),
\end{equation}
where $v_{\rm r+m}$ is the ``macroturbulence+rotation'' parameter
used in Paper II. 
These broadening parameters ($v_{\rm ip}$, $v_{\rm rt}$, 
and $v_{\rm mt}$) may be related to the more realistic quantities as
$v_{\rm ip} \simeq (c/R)/(2\sqrt{\ln 2})$ (2.00 km~s$^{-1}$ 
in the present case of $R\simeq 90000$), 
$v_{\rm rt} \simeq 0.94 v_{\rm e}\sin i$
($v_{\rm e}$ and $i$ are the equatorial rotation velocity and 
the inclination angle), and $v_{\rm mt} \simeq 0.42 \zeta_{\rm RT}$
($\zeta_{\rm RT}$: radial-tangential macroturbulence dispersion;
cf. Gray 2005), as explained in footnotes 10 and 12 of Takeda et al.
(2008).

Further, since we may reasonably postulate that the macroturbulence
velocity field in the solar atmosphere can be applied to
all of the three solar twin targets, we assume an analytical
form of the depth-dependent macroturbulence (in terms of 
$\tau_{5000}$, the optical depth at 5000~$\rm\AA$)
\begin{equation}
v_{\rm mt} (= 0.42 \zeta_{\rm RT}) = 1.60 - 0.11 \log \tau_{5000} 
- 0.19 \log \tau_{5000}^{2},
\end{equation} 
since $\zeta_{\rm RT}$ may be approximately expressed as 
$3.8 - 0.25 \log \tau_{5000} - 0.45 \log \tau_{5000}^{2}$
from figure 2 of Takeda (1995).

Now the procedure makes as follows.\\
(1) First, $v_{\rm M}$ is determined by applying
the profile-fitting method to a blend free Fe line.\\
(2) Then, $v_{\rm r+m}$ (macroturbulence plus rotation) is 
obtained by subtracting the instrumental broadening effect 
($v_{\rm ip}$ = 2.00 km~s$^{-1}$)
as $v_{\rm r+m} = \sqrt{v_{\rm M}^{2} - v_{\rm ip}^{2}}$.\\
(3) From the measured equivalent width, the mean-depth of
line formation ($\langle \log \tau_{5000} \rangle$) relevant
for this line can be computed (cf. subsection 5.1 in Takeda 1995),
which is sufficient to assign an appropriate value of 
$v_{\rm mt}$ to this line with the help of equation (2).\\
(4) Eventually, $v_{\rm rt}$ is evaluated as 
$v_{\rm rt} = \sqrt{v_{\rm r+m}^{2} - v_{\rm mt}^{2}}$.

\setcounter{figure}{2}
\begin{figure}
  \begin{center}
    \FigureFile(85mm,40mm){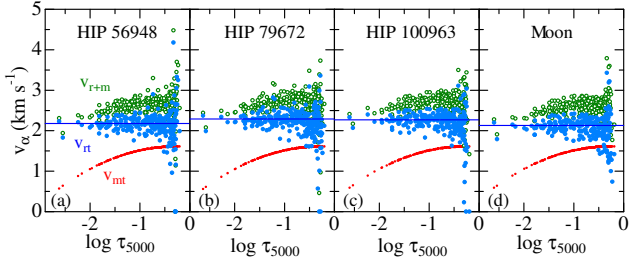}
  \end{center}
\caption{Line-broadening parameters (derived by the 
profile-fitting method applied to blend-free Fe lines)
plotted against the mean-depth of line formation
($\langle \log \tau_{5000} \rangle$), where green open 
circles, blue filled circles, and red dots represent
$v_{\rm r+m}$ (rotation+macroturbulence),  $v_{\rm mt}$ 
(macroturbulent velocity parameter, supposed to be 
depth-dependent as  $v_{\rm mt} = 1.60 - 0.11 \log \tau_{5000} 
- 0.19 \log \tau_{5000}^{2}$; cf. subsection 3.2) and  $v_{\rm rt}$
(rotational broadening parameter obtained as 
$(v_{\rm r+m}^{2} - v_{\rm mt}^{2})^{1/2}$), respectively.
The blue horizontal line indicates the average value of $v_{\rm rt}$
($\langle v_{\rm rt} \rangle$; cf. table 1) calculated for the lines 
of $\langle \log \tau_{5000} \rangle \le -0.7$.
(a) HIP~56948, (b) HIP~79672, (c) HIP~100963, and (d) Moon.
}
\end{figure}

While the detailed results of $v_{\rm r+m}$ and $v_{\rm rt}$ 
(along with the assigned $\langle \log \tau_{5000} \rangle$) 
for each line are presented in electronic table E2, 
these $v_{\rm r+m}$, $v_{\rm mt}$, and $v_{\rm rt}$ are plotted
against $\tau_{5000}$ in figure 3.
We can see from this figure that the depth-dependent tendency of 
$v_{\rm r+m}$ is almost removed in $v_{\rm rt}$ by subtracting
the effect of $v_{\rm mt}$. Finally, we obtained 
$\langle v_{\rm rt} \rangle$ as the parameter representing 
$v_{\rm e} \sin i$\footnote{Admittedly, we can not hope to relate
the ``exact'' value of $v_{\rm e} \sin i$ to 
$\langle v_{\rm rt} \rangle$ within the framework of such a 
rough modeling of line-broadening functions (all assumed to 
be the Gaussian form). However, we may reasonably expect that 
$\langle v_{\rm rt} \rangle$ is proportional 
to $v_{\rm e} \sin i$ with a factor not much different from
unity. This is sufficient for our present purpose, because
what we want to know is the ``differential'' characteristics
(i.e., the ratio of $\langle v_{\rm rt} \rangle$ between 
two stars is considered to be the ratio of actual 
$v_{\rm e} \sin i$). At any rate, it is encouraging that the 
resulting $\langle v_{\rm rt}\rangle$ value of 2.13 km~s$^{-1}$ for 
the Sun/Moon is quite close to the actual solar $v_{\rm e} \sin i$
value of 1.9 km~s$^{-1}$, by which we may regard that our 
approximation (suggesting $v_{\rm rt} \simeq 0.94 v_{\rm e}\sin i$) 
is not bad.} by averaging the $v_{\rm rt}$'s with 
$\langle \log \tau_{5000} \rangle \le -0.7$ (deep-forming lines
with $\langle \log \tau_{5000} \rangle \ge -0.7$ were not used
for the averaging because of the larger uncertainties due to 
the weakness of the line-strength), as given in table 1.

\subsection{Li Abundance}

The portion of the observed spectrum (6706.3--6709.3~$\rm\AA$) comprising
the Li~{\sc i} resonance doublet at $\sim$~6707.8~$\rm\AA$, along with 
Kurucz et al.'s (1984) solar flux spectrum, is shown in figure 4
(left panels). We can recognize from this figure that the strengths
of the Li line for HIP~79672 and HIP~100963 are markedly larger
than those for HIP~56948 and the Sun/Moon, classifying these four
into Li-strong and Li-weak groups.
As in Paper II, the Li abundance ($A_{\rm Li}$) was determined 
from the Li~{\sc i} doublet lines at $\sim 6707.8 \rm\AA$
in the same manner as described in Takeda and Kawanomoto (2005).
Namely, we first establish the LTE abundance ($A_{\rm Li}^{\rm LTE}$) 
by applying the method of synthetic profile fitting to the 
spectrum feature of Fe~{\sc i} +  Li~{\sc i} lines (see the right panels 
in figure 4). Then, the EW(Li~{\sc i} 6708) is inversely calculated
from such obtained $A_{\rm Li}^{\rm LTE}$. Finally, while taking
into account the non-LTE effect, $A_{\rm Li}^{\rm NLTE}$ is
calculated from EW(Li~{\sc i} 6708). The resulting 
$A_{\rm Li}^{\rm LTE}$ for each star is presented in table 1.
In all the four cases studied, the non-LTE correction 
$\Delta (\equiv A_{\rm Li}^{\rm NLTE} - A_{\rm Li}^{\rm LTE})$
turned out to be +0.07. The solar Li abundance of 0.91
derived in this study based on the spectrum of Moon (Subaru/HDS) 
is in excellent agreement with the result of 0.92 concluded by 
Takeda and Kawanomoto (2005) based on the spectrum of Moon 
(OAO/HIDES) as well as the solar flux spectrum (Kurucz et al. 1984).

\setcounter{figure}{3}
\begin{figure}
  \begin{center}
    \FigureFile(80mm,100mm){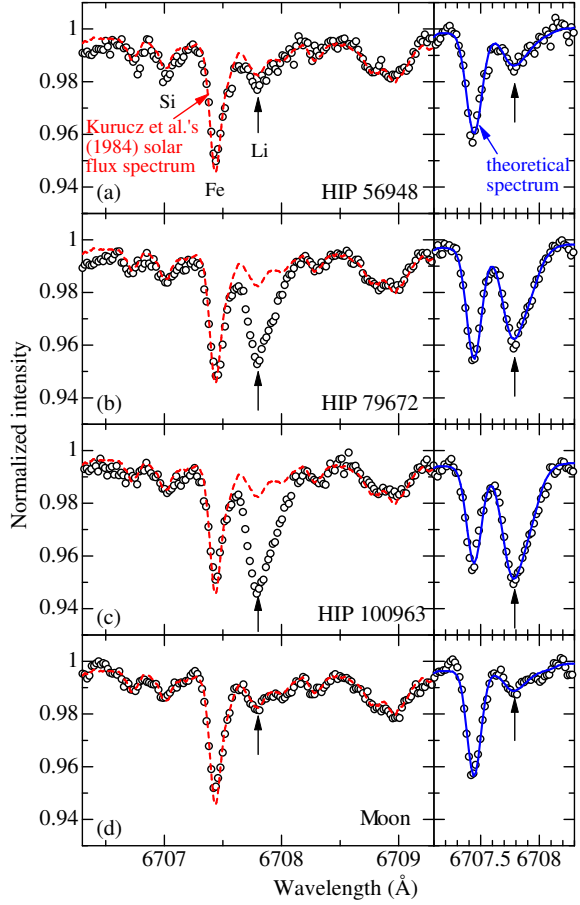}
  \end{center}
\caption{Left side: Observed spectra of target stars in the 
spectrum region comprising Li~{\sc i} 6707.8 line (open symbols) 
in comparison with the solar flux spectrum (red dashed line). 
Right side: Theoretically simulated spectrum in the Li~{\sc i} 6707.8 
line region (blue solid line) fitted with the observed stellar 
spectrum (open symbols). (a) HIP~56948, (b) HIP~79672, 
(c) HIP~100963, and (d) Moon.
}
\end{figure}

\subsection{Be Abundance}

The spectrum portion of 3129.5--3131.5~$\rm\AA$ comprising
Be~{\sc ii} lines at 3130.42~$\rm\AA$ and 3131.07~$\rm\AA$
is shown in figure 5, where each stellar spectrum is compared
with Kurucz et al.'s (1984) solar flux spectrum.
A glance of this figure suffices us to convince that Be line
features are essentially the same as the solar case for all stars.
According to the theoretical simulation shown in the lowest panel
of this figure, the agreement of $A_{\rm Be}$ with the solar value
appears to be very good, presumably to within $\sim 0.1$~dex
(though the uncertainty may be somewhat larger for HIP~56948 
where the spectrum quality is comparatively poor).
This fact clearly suggests that Be makes a clear distinction from Li 
(showing an appreciable difference from star to star) in spite of their 
rather similar characters comparatively easily destroyed in the stellar 
interior, at least for these solar twin stars are concerned. 
This result is consistent with what Randich et al. (2002) concluded
for early-G dwarf stars in open clusters.

\setcounter{figure}{4}
\begin{figure}
  \begin{center}
    \FigureFile(80mm,100mm){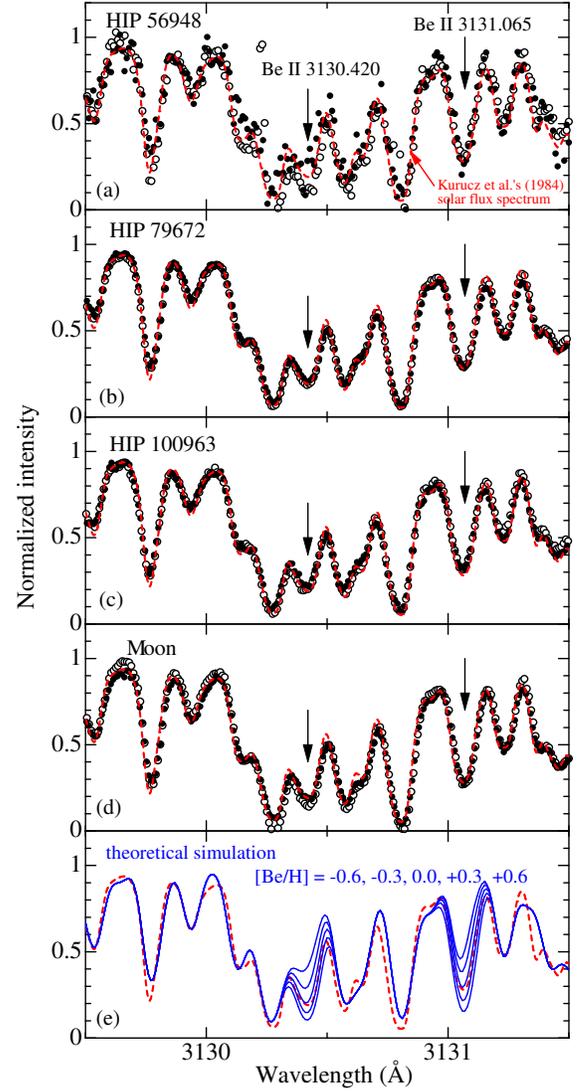}
  \end{center}
\caption{Observed spectra of target stars in the near UV region 
comprising two Be~{\sc ii} lines at 3130.42 and 3130.07 $\rm\AA$
(symbols: filled and open circles correspond the spectra
of different echelle orders 189 and 190, respectively), 
in comparison with Kurucz et al.'s (1984) solar flux spectrum 
(red dashed line). The continuum level of each spectrum has been 
so adjusted as to achieve a consistency between the stellar and 
the reference solar spectrum.(a) HIP~56948, (b) HIP~79672, 
(c) HIP~100963, and (d) Moon. In the lowest panel (e),
theoretically synthesized spectra for Sun (blue solid lines), 
are compared with the solar flux spectrum (red dashed line),
which were computed by using the atomic data given in Primas et al. 
(1997) with different Be abundances of [Be/H] = $-0.6$, $-0.3$, 0.0, 
+0.3, and +0.6. 
}
\end{figure}

\subsection{Differential Analysis}

Now that the ``standard'' atmospheric parameters were established
in subsection 3.1, we can derive the ``differential'' parameters
$\Delta p_{i-j}$ ($p$ is any of $T_{\rm eff}$, $\log g$, $v_{\rm t}$, 
and $A_{\rm Fe}$) of star $i$ relative to any other arbitrary 
comparison star $j$ by using the method described in Paper I,
where several practical quantities were defined such as (i)
the average of the direct solution, $\langle \Delta p_{ij} \rangle$ 
(average of $\Delta p_{i-j}$ and $-\Delta p_{j-i}$), (ii) the
intermediary solution via star $k$, $\langle \Delta p_{i(k)j} \rangle$
($\equiv \langle \Delta p_{ik} \rangle +\langle \Delta p_{ik} \rangle$),
and (iii) the average of the intermediary solution, 
$\langle \langle \Delta p_{i()j} \rangle \rangle$ (average of
$\langle \Delta p_{i(k)j} \rangle$ over various $k$).

Since we are interested in the parameter differences relative to
the Sun, we take $i = 1, 2, 3$ and $j =0$ (see table 1 for the
numbering of each star), and two intermediary stars can be
assigned for any pair (e.g., for the case of $i=1$ and $j=0$,
we can take $k = 2$ or $k = 3$).
The detailed results for HIP~56948 ($i=1$), HIP~79672 ($i=2$), 
HIP~100963 ($i=3$) are presented in tables 2, 3, and 4, respectively.
(Note that these three tables are formatted in the same manner as in
tables 2--9 of Paper I.) The values of $\langle \Delta p_{i0} \rangle$ 
(direct solution) and $\langle \langle \Delta p_{i()0} 
\rangle \rangle$ (average of the intermediary solution) are 
separately summarized in table 5, where the differences in $v_{\rm rt}$ 
and $A_{\rm Li}$ are also given. It is worth noting that 
the comparison of these two direct and intermediary solutions
may provide us with an opportunity of checking/estimating the 
accuracy of the results. 

\section{Discussion}

\subsection{Which Is the Best Solar Twin?}

When we compare the parameter differences (relative to the Sun) 
for HIP~79672 and HIP~100963 given in table 5 with those of Paper II 
(see table A.1 therein), we see a notable discrepancy in the values 
of $\Delta T_{\rm eff}$,
despite that other $\Delta \log g$, $\Delta v_{\rm t}$, and 
$\Delta A_{\rm Fe}$ are mostly in good agreement. Namely, while Paper II 
derived very small $\Delta T_{\rm eff}$ (+1.7~K/$-1.6$~K for 
HIP~79672/100963), the present results (+48.5~K/+38.2~K for the direct 
solution) indicate appreciably larger values by $\sim$~40--50~K than these.
Although the reason for this $\Delta T_{\rm eff}$ discrepancy is not clear, 
it would presumably be ascribed to the difference in the used EW data set. 
At any rate, because of the reasonable consistency between the direct 
and intermediary solutions (compare the values in the upper and lower 
rows in table 5) and the use of the spectrum data of much higher quality
(in terms of both the S/N ratio and the spectrum resolving power),
we would place larger weight in the present results.

By inspecting table 5, we can summarize as follows concerning 
the similarity or dissimilarity of these three program stars 
to the Sun in terms of each checkpoint. (In the discussion of the parameter
differences given below, we refer to the averaged values of the direct and 
intermediary solutions for convenience.) 
\begin{itemize}
\item
$\Delta T_{\rm eff}$: HIP~56948 is manifestly more solar like 
($\sim +10 \pm 10$~K) than other HIP~79672 and HIP~100963 
($\sim +40 \pm 10$~K).
\item
$\Delta \log g$: All three stars do not show appreciable differences 
from the solar value, if we consider the nominal uncertainty of 
$\sim 0.01$~dex.
\item
$\Delta v_{\rm t}$: HIP~79672 shows slightly higher $v_{\rm t}$ than 
the Sun by +0.03--0.04~km~s$^{-1}$, HIP~56948 and HIP~100963 are 
essentially solar.
\item
$\Delta A_{\rm Fe}$: Regarding the metallicity, HIP~100963 is almost
indiscernible from the Sun ($\ltsim 0.01$~dex) and HIP~56948 
($\sim$~+0.01--0.02 dex) is also near-solar (or very slightly 
metal-rich?), while HIP~79672 appears to be somewhat metal-rich 
($\sim$~+0.05~dex).
\item
$A_{\rm Li}$: HIP~79672 and HIP~100963 are markedly
overabundant in Li compared to the Sun by $\sim$~+0.7--0.8~dex
(by a factor of $\sim$~5--6), while the difference from $A_{\rm Li, \odot}$ 
is much milder for HIP~56946 (only $\sim +0.2$~dex or $\sim 60$\%).
\item
$\langle v_{\rm rt}\rangle$ (equivalent to $v_{\rm e}\sin i$):
HIP~56948 has almost the same projected rotational velocity
as the Sun, while HIP~79672 and HIP~100963 show slightly higher 
values by $\sim$~5--10\%.
\item
$A_{\rm Be}$: All three stars (HIP~56948, HIP~79672, and HIP~100963)
have essentially the same Be abundances as the Sun, which means that
Be does not conform to the behavior of Li showing a diversity. Consequently,
whichever mechanism changing the surface Li abundance of these solar twins
can not influence Be; e.g., if the variation of $A_{\rm Li}$ is caused
by an envelope mixing, it should not be so deep as to affect Be
(see also Randich et al. 2002).
\end{itemize}
Taking all these results into consideration, we can draw the following 
conclusions.

HIP~56948 is surely most similar to the Sun among these three 
stars, not only from the similarity of stellar parameters but also
from the viewpoint of surface Li abundance; it may thus deserve 
the name of ``closest ever solar twin.'' However, unlike the argument
of Mel\'{e}ndez and Ram\'{\i}rez (2007) who derived 
$\Delta A_{\rm Li} = -0.02 (\pm 0.13)$, since the atmospheric Li 
abundance of this star is marginally higher than the solar value 
by $\sim$~0.2~dex, we still can not call it a ``genuine'' solar twin. 
Another concern about this star is that its luminosity derived from
the Hipparcos parallax appears to somewhat larger than the solar
luminosity, which in effect makes the age older (cf. table 1).
We suspect that this inconsistency is attributed to the error in 
the parallax because HIP~56948 is comparatively distant.
The possibility that its actual $\pi$ is by $\sim$~10\% larger than 
the catalogued value may as well be considered, since the similarity
of $T_{\rm eff}$, $\log g$, and $A_{\rm Fe}$ should guarantee
the equality of $L$ (cf. Appendix A of Paper II).

Regarding HIP~79672 and HIP~100963, they have by $\sim 40$~K 
higher $T_{\rm eff}$ than $T_{\rm eff,\odot}$, by $\sim$~0.7--0.8~dex 
larger $A_{\rm Li}$ than $A_{\rm Li,\odot}$, and by $\sim$~5--10\% larger
$v_{\rm e}\sin i$ than $v_{\rm e}\sin i_{\odot}$. Apart from these 
considerable differences, the parameters of HIP~100963 quite resemble 
the solar values. 
Meanwhile, HIP~56948 shows other noticeable differences from the Sun 
with respect to $v_{\rm t}$ (by +0.03--0.04~km~s$^{-1}$) and 
$A_{\rm Fe}$ ($\sim$~+0.05~dex), which makes this star comparatively
lower ranked as a solar twin among the three. 

\subsection{What Controls Lithium? --- Roles of Rotation and $T_{eff}$}

Let us turn our attention to the question posed in section 1:
``why these solar twins show a diversity in the Li line strength
in spite of their similarity to one another?''
Is this attributed to the difference in the rotational velocity, 
as suggested in Paper II? 
According to tables 1 and 5, the values of $v_{\rm rt} (\simeq v_{\rm e}\sin i)$ 
for the Li-strong group (HIP~79672 and HIP~100963) are somewhat larger
by $\sim$~5--10\% than those for the Li-weak group (HIP~56948 and 
the Sun/Moon), which can also be visually recognized in figure 3.
Considering that $i=90^{\circ}$ for the Sun while $i$ is unknown 
for the three stars, we can assure that the equatorial rotational
velocities ($v_{\rm e}$) of Li-strong HIP~79672 and HIP~100963 are 
anyhow larger than the solar value ($v_{\rm e,\odot}$), which may be 
just favorable for the working hypothesis of Paper II.

Yet, we feel it still premature to conclude that the stellar rotation
is the only decisive factor to influence the surface Li abundance of
these solar twin stars. Inspecting the core features of Ca~{\sc ii} 
H and K lines of the program stars shown in figure 6, which are sensitive 
to the chromospheric activity closely related to the stellar 
rotation rate,\footnote{The strength of the Ca~{\sc ii} H+K 
core emission is known to roughly scale with the rotational rate 
(e.g., Noyes et al. 1984), though its exact relationship is still under 
discussion (see, e..g., Giampapa 2005). This means, for example, 
that a solar-type star rotating with $v_{\rm e} \sim 10$~km~s$^{-1}$ 
would show a stronger core emission by several times than the Sun.}
we see in any of these stars almost no appreciable differences in comparison 
to the solar H and K cores of Kurucz et al.'s (1984) solar spectrum.
(Actually, the largest difference relative to the Sun among the four 
is seen in the ``Moon'' spectrum, which is presumably due to the difference 
in the solar activity phase, because the year of 2008 corresponds to 
the almost minimum activity.) 
Besides, according to Giampapa (2005; cf. figure 36 therein), the Li-strong 
HIP~79672 exhibits the stellar activity (inferred from Ca~{\sc ii} H and K 
line cores; amplitude of $\sim 10$\% with period of 8--9 years) quite 
similar to the solar activity. We thus consider that ``substantially'' large
difference in the rotational rate is not very likely between Li-strong
(HIP~79672 and HIP~100963) and Li-weak (HIP~56948 and the Sun) groups.

\setcounter{figure}{5}
\begin{figure}
  \begin{center}
    \FigureFile(70mm,100mm){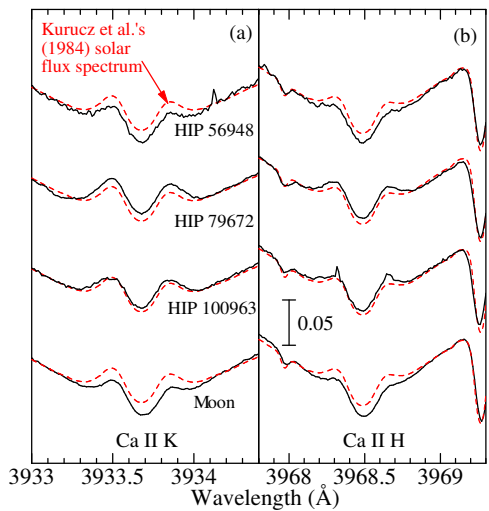}
  \end{center}
\caption{Emission features in (a) Ca~{\sc ii} K and (b) Ca~{\sc ii} H 
resonance line cores of the program stars (solid lines), compared 
with Kurucz et al.'s (1984) solar flux spectrum (red dashed line). 
The continuum level of each spectrum (with a vertical offset of 1.0 
relative to the adjacent one) has been appropriately adjusted 
so that the stellar spectrum matches the reference solar spectrum 
at $\Delta \lambda \sim \pm 0.5 \rm\AA$. 
(a) HIP~56948, (b) HIP~79672, (c) HIP~100963, and (d) Moon.

}
\end{figure}

We rather suspect that the difference in $T_{\rm eff}$ may also play 
a significant role in producing this Li-diversity. 
As reported in Paper II, there is a slanted ``lower boundary line'' 
at $+100$~K~$\gtsim \Delta T_{\rm eff} \gtsim$~0~K 
in the $A_{\rm Li}$ vs. $\Delta T_{\rm eff}$ diagram, below which  
no stars are found (cf. figure 9 therein).
Interestingly, when we plot HIP~79672 and HIP~100963 (both have
$\Delta T_{\rm eff} \sim +50$~K and $A_{\rm Li} \sim 1.6$) 
on this diagram, they almost fall on this boundary line, 
$A_{\rm Li} \simeq 1 + 1(\Delta T_{\rm eff}/100\,{\rm K})$,
which we regard to be a significant fact being worth attention.

That is, as speculated in Paper II, we consider that the diversity of 
$A_{\rm Li}$ (at a given $T_{\rm eff}$) is due to the difference in 
the rotational velocity (i.e., slower rotators tend to show lower 
Li abundances presumably caused by an enhanced envelope-mixing). 
On the other hand, since the existence of steeply-slanted 
lower boundary of $A_{\rm Li}$ means that every slow rotators should 
settle on this boundary line, the $A_{\rm Li}$ values of such slow 
rotators would naturally show the marked $T_{\rm eff}$-dependence of 
$dA_{\rm Li}/d (T_{\rm eff}/100\,{\rm K}) \sim 1$. This scenario may 
reasonably explain (at least to an order of magnitude) the difference 
in $A_{\rm Li}$ by $\sim$~0.6--0.7 dex between HIP~79672/HIP~100963 
and Sun/HIP~79672, all showing superficially slow rotation as the Sun, 
while $T_{\rm eff}$ for the former being slightly higher than the 
latter by $\sim$~40--50~K.
Consequently, the Li-strong nature of HIP~79672 and HIP~100963 may be 
naturally explained by the fact that they belong to the ``boundary-line 
stars'' (which seem to have higher possibilities of hosting planets; cf. 
subsection 5.1 in Paper II). In this sense, we would suggest that 
$T_{\rm eff}$ is another significant factor (along with rotational 
velocity $v_{\rm e}$) in controlling the lithium abundances of 
solar twins, especially for slowly-rotating ones.

Anyway, this is nothing but a phenomenological explanation, and
a number of tasks are still left until the real physical mechanism 
involved in determining the surface Li abundance is clarified. 
Hence, investigations (especially on the theoretical side) on the 
inter-relations between rotation, $T_{\rm eff}$, and $A_{\rm Li}$ in 
solar-analog stars are desirably awaited, so that the confronted 
problems could be settled:\\
--- Why does such a slanted lower boundary exist in the 
$A_{\rm Li}$ vs. $T_{\rm eff}$ diagram of solar-analog stars,
below which stars do not exist (``forbidden zone'')? Any
$T_{\rm eff}$-sensitive physical mechanism is acting so as to
suppress the further Li depletion? \\
--- $A_{\rm Li}$ appears to be positively correlated with 
both $v_{\rm e}\sin i$ and $T_{\rm eff}$. What does this mean? 
These two factors happen to act independently on $A_{\rm Li}$ in the
same direction? Or this is nothing but a superficial effect caused 
by a tight relationship between $v_{\rm e}\sin i$ and $T_{\rm eff}$?\\
--- Whichever $v_{\rm e}\sin i$ or $T_{\rm eff}$ may be the relevant key, 
the physical mechanism working in the envelope of these solar 
analog stars must satisfy the condition of changing the surface 
Li without affecting Be. What kind of process is that?\\
--- Finally, from the observational side, the number of well-studied 
solar twins/analogs is still so insufficient as to clearly reveal the 
behavior of Li in Sun-like stars. If we could considerably increase 
the number of the sample stars (e.g., $\sim 10^3$ or even more), 
it would surely give us a new insight to this field (in addition, 
a nearly-perfect solar twin might as well be detected). Besides, 
given that field solar-type stars are diverse in their age (cf. 
figure 10 in Paper II), intensively studying the early-G dwarfs 
in old solar-age clusters (e.g., M~67) would also be beneficial for 
disentangling the roles of various stellar parameters on this Li problem.

\section{Conclusion}

An intensive spectroscopic study based on the high-quality spectra 
obtained with Subaru/HDS was performed for HIP~56948, HIP~79672, and 
HIP~100963 (along with the Sun/Moon as the reference standard), 
known to be the representative solar twins, in order to
(1) clarify which of the three most resembles the Sun 
by precisely establishing the various differential parameters 
relative to the Sun and (2) investigate the reason why appreciable 
differences in the surface Li abundance are observed for these 
superficially very similar stars.

The standard atmospheric parameters ($T_{\rm eff}^{\rm std}$, 
$\log g^{\rm std}$, $v_{\rm t}^{\rm std}$, and \{Fe/H\}$^{\rm std}$) 
were first evaluated by using the equivalent widths of Fe~{\sc i} and 
Fe~{\sc ii} lines, the rotational velocity parameter ($v_{\rm rt}$; 
which is nearly equivalent to $v_{\rm e}\sin i$) was derived from
the line-profile width by eliminating the effect of the macroturbulence,
and the lithium abundance ($A_{\rm Li}$) was determined from the 
Li~{\sc i} doublet line at $\sim$~6707.8~$\rm\AA$. 
Further, the differences of atmospheric parameters 
($\Delta T_{\rm eff}$, $\Delta \log g$, $\Delta v_{\rm t}$, and 
$\Delta A_{\rm Fe}$) relative to the Sun were established
by using the method of precision differential analysis (Paper I).

While we could confirm that HIP~79672/100963 have appreciably higher Li 
content by a factor of $\sim$~5--6  as compared to Sun/HIP~56948,
the Be abundances for all the program stars (HIP~56948/79672/100963)
turned out to be essentially the same as the solar value, which 
indicates that Be is not affected by any mechanism causing the variation 
of Li. 

We found that HIP~56948 is most similar to the Sun among the three, 
not only from the similarity of stellar parameters (including rotation) 
but also from the weakness of the Li line (however, $A_{\rm Li}$ 
for this star is still slightly larger than $A_{\rm Li,\odot}$ by 
$\sim 0.2$~dex; i.e., not perfectly the same). It may thus deserve 
the name of ``closest ever solar twin.'' Meanwhile, some remarkable
differences from the solar parameters are recognized in HIP~79672 and 
HIP~100963, which show somewhat higher $T_{\rm eff}$ (by $\sim 40$~K)
considerably larger $A_{\rm Li}$ (by $\sim$~0.7--0.8~dex)
and slightly higher rotational velocity (by $\sim$~5--10\%). 

We can see a tendency that the Li-strong HIP~79672 and HIP~100963 
have somewhat larger rotational velocity by $\sim$~5--10\% than  
Li-weak HIP~56948 and the Sun, which is consistent with the suggestion 
of Paper II that $A_{\rm Li}$ is closely correlated with the stellar 
rotational velocity. However, it does not seem very likely that a
substantial difference exists in the rotational velocity between these
two groups, because no essential differences are seen in their 
chromospheric activities (sensitive to stellar rotation) inferred 
from Ca~{\sc ii} H+K line cores. 
We rather suspect that the overabundance of Li in HIP~79672 and 
HIP~100963 (by $\sim$~0.6--0.7~dex) is attributed to the difference 
in $T_{\rm eff}$ (by $\sim$~+50~K) relative to the Sun, since
these two stars fall on the $T_{\rm eff}$-sensitive slanted lower 
boundary in the $A_{\rm Li}$ vs. $T_{\rm eff}$ distribution as reported
in Paper II for solar analog stars. However, it is not clear whether 
and how these two factors (rotational velocity and effective temperature)
are mutually related in affecting the surface Li abundance, which remains
to be further clarified.



\onecolumn

\setcounter{table}{0}
\setlength{\tabcolsep}{3pt}
\begin{table}[h]
\scriptsize
\caption{Target stars and their parameters}
\begin{center}
\begin{tabular}{ccccccccccrcccrc}\hline\hline
Star & HIP &  Name & Sp. type & $T_{\rm eff}^{\rm std}$ & 
$\log g^{\rm std}$ & $v_{\rm t}^{\rm std}$ & \{Fe/H\}$^{\rm std}$ & 
$\pi$ & $\sigma/\pi$ & $\log L$ & $M$ & $\log age$ & 
$\langle v_{\rm rt}\rangle$ & $EW_{6708}$ & $A_{\rm Li}^{\rm NLTE}$ \\
No. & number  &  &  & (K) & (cm~s$^{-1}$) & (km~s$^{-1}$) &
 (dex) & (mas) &  & ($L_{\odot}$) & ($M_{\odot}$) & (yr) & (km~s$^{-1}$) &
 (m$\rm\AA$) & \\
\hline
1 & ~56948  & $\cdots$ & G5 & 5747.9 & 4.409 & 0.93 & $-0.016$ & 15.0 & 0.05 & +0.13 & 0.98 & 9.90 &
 2.18 & 3.6 & 1.13\\
2 & ~79672  & 18~Sco & G1~V  & 5771.7 & 4.397 & 0.97 & +0.011 & 71.3 & 0.01 & +0.05 & 1.01 & 9.70 &
 2.29 & 9.7 & 1.60\\
3 & 100963 & $\cdots$ & G5 & 5760.0 & 4.411 & 0.93 & $-0.040$ & 35.4 & 0.02 & +0.02 & 1.00 & 9.66 &
 2.27 & 12.1 & 1.68\\
\hline
0 & $\cdots$ & Sun/Moon & G2~V & 5737.1 & 4.420 & 0.95 & $-0.036$ & $\cdots$ & $\cdots$ & (0.00) & (1.00) & (9.66) &
 2.13 & 2.2 & 0.91\\
\hline
\end{tabular}
\end{center}
Notes:\\ 
In columns 5--8 are listed the $T_{\rm eff}^{\rm std}$ 
(effective temperature), $\log g^{\rm std}$ (logarithmic surface gravity), 
$v_{\rm t}^{\rm std}$ (microturbulence), and  
\{Fe/H\}$^{\rm std}$ ($\equiv A_{\rm Fe}^{\rm std} - 7.50$; metallicity),
which are the ``standard'' atmospheric parameters spectroscopically 
determined based on the selected Fe~{\sc i} and Fe~{\sc ii} lines 
(see subsubsection 3.1.1 of Paper II).
Columns 9 and 10 present the Hipparcos parallaxes (ESA 1997) and 
their relative errors, while the values of luminosity, mass, and age 
are given in columns 11-13, which were derived from 
the positions on the theoretical HR diagram with the help of stellar 
evolutionary tracks (see subsection 3.3 of Paper II). 
The $\langle v_{\rm rt}\rangle$ (column 14) is the rotational 
broadening parameter (nearly equivalent to the projected rotational 
velocity $v_{\rm e}\sin i$) determined from the widths of a number of
Fe~{\sc i} and Fe~{\sc ii} lines (cf. subsection 3.2). The equivalent width 
of the Li~{\sc i}~6707.8 doublet and the logarithmic abundance of Li (including 
the relevant non-LTE correction of +0.07 dex for all four stars) in 
the usual normalization of $A_{\rm H} = 12$ are presented in the 
last columns 15--16.\\
\end{table}

\setcounter{table}{1}
\setlength{\tabcolsep}{3pt}
\begin{table}[h]
\small
\caption{Differential analysis of HIP 56948 relative to the Sun.}
\begin{center}
\begin{tabular}{crrrrrrrrrrrrr}\hline\hline
\multicolumn{14}{c}{[direct analysis]}\\
 & $\Delta T$ & $\Delta\log g$ & $\Delta v_{\rm t}$ & $\Delta A$ &
   $\epsilon_{T}$ & $\epsilon_{g}$ & $\epsilon_{v}$ & 
   $\epsilon_{A_{1}}$ & $\epsilon_{A_{2}}$ & 
   $\sigma_{A_{1}}$ & $\sigma_{A_{2}}$ & 
   $N_{\rm 1}$ & $N_{\rm 2}$ \\ 
\hline
$  056948 - {\rm Sun} $ & $ +4.3$ & $-0.015$ & $-0.02$ &  $+0.015$ &
  5.0  & 0.010 & 0.04 & 0.007 & 0.008 &
 0.021 & 0.021 & 196  & 18 \\ \smallskip
$-({\rm Sun} - 056948)$ & $ +1.3$ & $-0.027$ & $-0.01$ &  $+0.012$ &
  5.0  & 0.010 & 0.04 & 0.006 & 0.007 &
 0.020 & 0.018 & 189  & 16 \\
$\langle 056948 - {\rm Sun} \rangle$ & $ +2.8$ & $-0.021$ & $-0.01$ &  $+0.013$ &
\multicolumn{9}{c}{} \\
\hline
\multicolumn{14}{c}{[indirect analysis]}\\
   & $\Delta T$ & $\Delta\log g$ & $\Delta v_{\rm t}$ & $\Delta A$ &
   $\sigma_{T}$ & $\sigma_{g}$ & $\sigma_{v}$ & $\sigma_{A}$ &
   \multicolumn{5}{c}{} \\
\hline \smallskip
$\langle\langle 056948 - {\rm Sun} \rangle\rangle$ & 
$+17.0$ & $+0.006$ & $-0.01$ & $+0.020$ & 
  3.8 & 0.012 & 0.01 & 0.004 & \multicolumn{5}{c}{} \\
(via 079672) & $+20.7$ & $+0.018$ & $-0.01$ & $+0.024$ & \multicolumn{9}{c}{}\\
(via 100963) & $+13.2$ & $-0.006$ & $+0.00$ & $+0.016$ & 
\multicolumn{9}{c}{[differences of standard parameters]} \\
 &  &  &  &  &
 & & \multicolumn{7}{l}{$\Delta T =       5747.9 - 5737.1  =  +10.8$} \\
 &  &  &  &  &
 & & \multicolumn{7}{l}{$\Delta\log g =    4.409 -  4.420  = -0.011$} \\
 &  &  &  &  &
 & & \multicolumn{7}{l}{$\Delta v_{\rm t} = 0.93 -   0.95  =  -0.02$} \\
 &  &  &  &  &
 & & \multicolumn{7}{l}{$\Delta A =        -0.016 - (-0.036)  = +0.020$} \\
\hline
\end{tabular}
\end{center}
Notes. \\
The results of differential analyses for the case of ($i=1$ and $j=0$). 
The brief description of the data in the table is given below
while Paper I should be consulted for more detailed explanations.
(Note that the effective temperature $T_{\rm eff}$ and the Fe abundance 
$A_{\rm Fe}$ are abbreviated as $T$ as $A$, respectively, 
in this table 2 along with the following tables 3 and 4.)\\
\\
(Direct analysis:)\\
--- 1st row: the results of ($\Delta T_{i-j}$, $\Delta \log g_{i-j}$, 
$\Delta v_{i-j}$, $\Delta A_{i-j}$), the possible errors ($\epsilon_{T}$, 
$\epsilon_{g}$, and $\epsilon_{v}$) involved in these solutions (estimated 
by the procedure described in subsection 5.2 of Takeda et al. 2002), the 
root-mean-square errors ($\epsilon_{A_{1}}$, $\epsilon_{A_{2}}$) on 
the differential abundances ($\Delta A_{1,i-j}$ and $\Delta A_{2,i-j}$) 
from Fe~{\sc i} and Fe~{\sc ii} lines corresponding to these ambiguities 
in atmospheric parameters, the standard deviations ($\sigma_{A_{1}}$ and 
$\sigma_{A_{2}}$) around the means of $\Delta A_{1,i-j}$ and 
$\Delta A_{2,i-j}$, and the numbers ($N_{1}$ and $N_{2}$) of the used 
Fe~{\sc i} and Fe~{\sc ii} lines. \\
--- 2nd row: the same as the 1st row, but for the inverse case of $j-i$;
i.e., presented are the parameter differences of ($-\Delta T_{j-i}$, 
$-\Delta \log g_{j-i}$, $-\Delta v_{j-i}$, and $-\Delta A_{j-i}$)
and the corresponding errors ($\epsilon_{T}$, $\epsilon_{g}$, 
$\epsilon_{v}$, $\epsilon_{A_{1}}$, and $\epsilon_{A_{2}}$).\\
--- 3rd row: averaged solutions of the parameter differences given 
in the 1st and 2nd rows; i.e., $\langle \Delta T_{ij} \rangle$, 
$\langle \Delta \log g_{ij} \rangle$, $\langle \Delta v_{ij} \rangle$,
and $\langle \Delta A_{ij} \rangle$.\\
\\ 
(Indirect analysis:)\\
--- The first row gives $\langle\langle \Delta p_{i()j} \rangle\rangle$ 
[equation (15) in Paper I] and $\langle\langle \sigma_{\Delta p,i()j} \rangle\rangle$ 
[equation (16) in Paper I], where $p$ denotes each of $T$, $\log g$, $v$, and $A$. \\
--- Meanwhile, in the following two rows are presented the individual 
$\langle \Delta p_{i(k)j} \rangle$ values [equation (14) in Paper I] for each 
intermediary star $k$ (from which the 
$\langle\langle \Delta p_{i()j} \rangle\rangle$ 
and $\langle\langle \sigma_{\Delta p,i()j} \rangle\rangle$ values 
in the first row were computed).\\
\\
(Inset in the lower-right space:) \\
--- The ($i-j$) differences of the standard parameters 
($T$, $\log g$, $v$, and $A$) given in table 1 . 
\end{table}

\setcounter{table}{2}
\setlength{\tabcolsep}{3pt}
\begin{table}[h]
\small
\caption{Differential analysis of HIP 79672 relative to the Sun.$^{*}$}
\begin{center}
\begin{tabular}{crrrrrrrrrrrrr}\hline\hline
\multicolumn{14}{c}{[direct analysis]}\\
 & $\Delta T$ & $\Delta\log g$ & $\Delta v_{\rm t}$ & $\Delta A$ &
   $\epsilon_{T}$ & $\epsilon_{g}$ & $\epsilon_{v}$ & 
   $\epsilon_{A_{1}}$ & $\epsilon_{A_{2}}$ & 
   $\sigma_{A_{1}}$ & $\sigma_{A_{2}}$ & 
   $N_{\rm 1}$ & $N_{\rm 2}$ \\ 
\hline
$  079672 - {\rm Sun} $ & $+48.9$ & $+0.008$ & $+0.03$ &  $+0.056$ &
  5.0  & 0.010 & 0.03 & 0.006 & 0.006 &
 0.018 & 0.016 & 194  & 17 \\ \smallskip
$-({\rm Sun} - 079672)$ & $+48.1$ & $+0.009$ & $+0.03$ &  $+0.053$ &
  5.0  & 0.010 & 0.03 & 0.006 & 0.006 &
 0.019 & 0.017 & 193  & 17 \\
$\langle 079672 - {\rm Sun} \rangle$ & $+48.5$ & $+0.008$ & $+0.03$ &  $+0.054$ &
\multicolumn{9}{c}{} \\
\hline
\multicolumn{14}{c}{[indirect analysis]}\\
   & $\Delta T$ & $\Delta\log g$ & $\Delta v_{\rm t}$ & $\Delta A$ &
   $\sigma_{T}$ & $\sigma_{g}$ & $\sigma_{v}$ & $\sigma_{A}$ &
   \multicolumn{5}{c}{} \\
\hline \smallskip
$\langle\langle 079672 - {\rm Sun} \rangle\rangle$ & 
$+39.3$ & $-0.013$ & $+0.04$ & $+0.048$ & 
  8.8 & 0.018 & 0.00 & 0.004 & \multicolumn{5}{c}{} \\
(via 056948) & $+30.6$ & $-0.031$ & $+0.03$ & $+0.044$ & \multicolumn{9}{c}{}\\
(via 100963) & $+48.1$ & $+0.005$ & $+0.04$ & $+0.052$ & 
\multicolumn{9}{c}{[differences of standard parameters]} \\
 &  &  &  &  &
 & & \multicolumn{7}{l}{$\Delta T =       5771.7 - 5737.1  =  +34.6$} \\
 &  &  &  &  &
 & & \multicolumn{7}{l}{$\Delta\log g =    4.397 -  4.420  = -0.023$} \\
 &  &  &  &  &
 & & \multicolumn{7}{l}{$\Delta v_{\rm t} = 0.97 -   0.95  =  +0.02$} \\
 &  &  &  &  &
 & & \multicolumn{7}{l}{$\Delta A =        0.011 - (-0.036)  = +0.047$} \\
\hline
\end{tabular}
\end{center}
$^{*}$The results of differential analyses for the case of ($i=2$ and $j=0$).
See the notes in table 2 for the details. 
\end{table}

\setcounter{table}{3}
\setlength{\tabcolsep}{3pt}
\begin{table}[h]
\small
\caption{Differential analysis of HIP 100963 relative to the Sun.$^{*}$}
\begin{center}
\begin{tabular}{crrrrrrrrrrrrr}\hline\hline
\multicolumn{14}{c}{[direct analysis]}\\
 & $\Delta T$ & $\Delta\log g$ & $\Delta v_{\rm t}$ & $\Delta A$ &
   $\epsilon_{T}$ & $\epsilon_{g}$ & $\epsilon_{v}$ & 
   $\epsilon_{A_{1}}$ & $\epsilon_{A_{2}}$ & 
   $\sigma_{A_{1}}$ & $\sigma_{A_{2}}$ & 
   $N_{\rm 1}$ & $N_{\rm 2}$ \\ 
\hline
$  100963 - {\rm Sun} $ & $+37.8$ & $+0.021$ & $+0.01$ &  $+0.003$ &
  0.0  & 0.010 & 0.04 & 0.005 & 0.007 &
 0.019 & 0.013 & 194  & 18 \\ \smallskip
$-({\rm Sun} - 100963)$ & $+38.6$ & $+0.026$ & $-0.01$ &  $+0.005$ &
  0.0  & 0.010 & 0.03 & 0.005 & 0.006 &
 0.018 & 0.014 & 188  & 17 \\
$\langle 100963 - {\rm Sun} \rangle$ & $+38.2$ & $+0.023$ & $+0.00$ &  $+0.004$ &
\multicolumn{9}{c}{} \\
\hline
\multicolumn{14}{c}{[indirect analysis]}\\
   & $\Delta T$ & $\Delta\log g$ & $\Delta v_{\rm t}$ & $\Delta A$ &
   $\sigma_{T}$ & $\sigma_{g}$ & $\sigma_{v}$ & $\sigma_{A}$ &
   \multicolumn{5}{c}{} \\
\hline \smallskip
$\langle\langle 100963 - {\rm Sun} \rangle\rangle$ & 
$+33.2$ & $+0.018$ & $-0.01$ & $+0.004$ & 
  5.4 & 0.009 & 0.00 & 0.002 & \multicolumn{5}{c}{} \\
(via 056948) & $+27.8$ & $+0.009$ & $-0.01$ & $+0.001$ & \multicolumn{9}{c}{}\\
(via 079672) & $+38.6$ & $+0.027$ & $-0.01$ & $+0.006$ & 
\multicolumn{9}{c}{[differences of standard parameters]} \\
 &  &  &  &  &
 & & \multicolumn{7}{l}{$\Delta T =       5760.0 - 5737.1  =  +22.9$} \\
 &  &  &  &  &
 & & \multicolumn{7}{l}{$\Delta\log g =    4.411 -  4.420  = -0.009$} \\
 &  &  &  &  &
 & & \multicolumn{7}{l}{$\Delta v_{\rm t} = 0.93 -   0.95  =  -0.02$} \\
 &  &  &  &  &
 & & \multicolumn{7}{l}{$\Delta A =        -0.040 - (-0.036)  = -0.004$} \\
\hline
\end{tabular}
\end{center}
$^{*}$The results of differential analyses for the case of ($i=3$ and $j=0$).
See the notes in table 2 for the details. 
\end{table}

\setcounter{table}{4}
\setlength{\tabcolsep}{3pt}
\begin{table}[h]
\small
\caption{Summary of differential parameters relative to the Sun}
\begin{center}
\begin{tabular}{crccccc}\hline\hline
Star & 
$\Delta T_{\rm eff}$ & $\Delta\log g$ & $\Delta v_{\rm t}$ & 
$\Delta A_{\rm Fe}$ & $\Delta A_{\rm Li}$ &
$\langle v_{\rm rt}\rangle/\langle v_{\rm rt}^{\odot}\rangle $ \\
 & (K) & (dex) & (km~s$^{-1}$) & (dex) & (dex) &  \\ 
\hline
HIP 56948  & $ +2.8$ & $-0.021$ & $-0.01$ & $+0.013$ & +0.22 & 1.02 \\
           & $+17.0$ & $+0.006$ & $-0.01$ & $+0.020$ &       &      \\
\hline                                               
HIP 79672  & $+48.5$ & $+0.008$ & $+0.03$ & $+0.054$ & +0.69 & 1.08 \\
           & $+39.3$ & $-0.013$ & $+0.04$ & $+0.048$ &       &      \\
\hline                                               
HIP 100963 & $+38.2$ & $+0.023$ & $+0.00$ & $+0.004$ & +0.77 & 1.07 \\
           & $+33.2$ & $+0.018$ & $-0.01$ & $+0.004$ &       &      \\
\hline
\end{tabular}
\end{center}
Notes:\\ 
Columns 2 through 5 present the ``star$-$Sun'' differences for each 
of the parameters ($T_{\rm eff}$, $\log g$, $v_{\rm t}$, and $A_{\rm Fe}$;
final averaged results extracted from tables 2--4) resulting from the 
differential analysis based on the method of Paper I, where the values
in the upper and lower row correspond to the direct solution 
($\langle \Delta p_{i0} \rangle$; cf. section 3 in Paper I) and 
the solution obtained via intermediary stars 
($\langle \langle \Delta p_{i()0} \rangle \rangle$; cf. section 4 in
Paper I), respectively.  Given in columns 6 and 7 are the differential
Li abundance relative to the Sun and the ``star/Sun'' ratio of 
$\langle v_{\rm rt} \rangle$
(rotational broadening parameter), respectively, which were simply
obtained from the data in table 1.
\end{table}

\newpage


\end{document}